\documentclass[12pt]{iopart}
\pdfoutput=1

\usepackage{setstack}
\usepackage{graphicx}
\usepackage{epstopdf}
\usepackage{amssymb}
\usepackage{default}

\newcommand{\be}{\begin{equation}}
\newcommand{\bea}{\begin{eqnarray}}
\newcommand{\bc}{\begin{center}}            
\newcommand{\ee}{\end{equation}}
\newcommand{\eea}{\end{eqnarray}}
\newcommand{\ec}{\end{center}}

\newcommand{\baa}{\begin{eqnarray*}}
\newcommand{\eaa}{\end{eqnarray*}}
\begin{document}
\title{Feynman-Smoluchowski engine at high temperatures and the role of constraints}

\author{Varinder Singh and Ramandeep S. Johal}

\address{Department of Physical Sciences, \\ 
Indian Institute of Science Education and Research Mohali,
Sector 81, S.A.S. Nagar, Manauli PO 140306, Punjab, India}
\ead{varindersingh@iisermohali.ac.in and rsjohal@iisermohali.ac.in}

\begin{abstract}
Feynman's ratchet and pawl is a paradigmatic model for energy conversion 
using thermal fluctuations in the mesoscopic regime. Here, 
we optimize the power output of the ratchet as a heat engine
in the high temperatures limit, and  
derive the universality of efficiency at maximum power up to second order,
using a non-linear approximation. 
On the other hand, the linear model may be optimized by constraining the internal energy
scales in different ways. It is shown that simple constraints lead
to  well-known expressions of thermal efficiency in finite-time thermodynamics. 
Thereby, the constrained ratchet, in the linear regime, has been 
mapped to an effective finite-time thermodynamic model.
\end{abstract} 
\maketitle
\section{Introduction}
Feynman-Smoluchowski (FS) ratchet \cite{Feynman,Smoluchowski} has motivated the modeling
of brownian or molecular motors \cite{Julicher,Parrondoreview,Asfaw,BaoQuan,ZhangLin,Lindner,
Astumian1997,Hanggi2009,Patrick2016} and sharpened the understanding of thought experiments
like Maxwell's demon \cite{LeffRex,Broeckdemon}. Subsequently, various analogs and generalizations
\cite{Magnasco1998,Brillouin1950,Jarzynski,ShengYang,Velasco,Tu2008,GJ2015,LongLiu2015,
VarinderJohal,Montiel2017,Apertet2014}, have been studied in literature. The ratchet is designed
to rectify thermal fluctuations across a mechanical link whose two asymmetric ends experience
different fluctuations due to their being embedded in different (hot and cold) baths.  
The processes of heat and work transfer are assumed to occur at finite rates, thus 
generating a finite output power. Feynman's analysis \cite{Feynman} concluded that the device 
could operate with reversible efficiency in the quasi-static limit  
which implies a vanishing output power. Based on this analysis, we shall also assume a 
strong coupling between the fluxes, i.e., there is no heat leakage between 
the heat baths (see \cite{Parrondo,Apertet2014,Park2017} for contrasting views). 
 
In the finite-power regime, one may extract maximum power by tuning the system's 
internal energy scales to appropriate values \cite{Velasco,Tu2008,VarinderJohal}. 
A related quantity of interest, that also places the FS system in a broader 
thermodynamic context, is the efficiency at maximum power (EMP).
It shares a universal property with many other finite-time models 
\cite{Esposito2010,Curzon1975,Esposito}, viz. for small differences in bath temperatures,
EMP behaves as ${\eta_c}/{2} + {\eta_c^2}/{8} + O[\eta_{c}^3]$, 
where $\eta_c = 1 - T_2/T_1$ is the Carnot bound, with $T_2 (T_1)$ as the cold (hot) bath temperatures.  
The first-order term can be explained using the strong-coupling assumption 
within linear irreversible thermodynamics \cite{Broeck2005}, while  
the second-order term is beyond linear response, and has been related to a certain
symmetry property in the model \cite{Lindenberg2009,JohalRai2016}. 

In this paper, we focus on the performance of FS ratchet  
at maximum power in the regime where thermal energy of a bath 
is much higher than the internal energy scale excited by the bath. 
We highlight new features of the device in this regime, not discussed 
earlier in literature. We note that it is not possible to optimize 
power---simultaneously over both internal scales---within the linear regime. 
However, a two-parameter optimization is possible if one extends the 
operational domain to non-linear approximation. Interestingly, one is able 
to then recover EMP that retains the same universality up to second order
as for the EMP of the original problem, Eq. (\ref{effTu}) below. 
We then impose some simple constraints over the internal energy scales, 
such that optimization of power over a single parameter can be performed 
using the linear model. These constrained optimization problems
yield some well-known forms of EMP found in other finite-time models. 
Moreover, under each of these constraints, it is possible
to give an effective finite-time thermodynamic model for the FS engine. 

The plan of the paper is as follows. In Sec. II, we briefly describe 
the model of FS engine and discuss its optimal performance. In Sec. III, 
two-parameter optimization of ratchet engine in high temperatures limit is
discussed. Sec. IV is devoted to optimization of the ratchet 
in linear regime, subject to constraints.
In Sec. V, FS engine is mapped to effective thermodynamic 
models depending on the constraints used in the previous section. Sec. VI is
devoted to a discussion of the results, with concluding remarks.

%
\section{Feynman's ratchet and pawl model}
Feynman's model \cite{Feynman} consists 
of a vane, immersed in a hot reservoir at temperature $T_1$,  
and connected through an axle with a ratchet in contact with a
cold reservoir at $T_2$. In the center of the axle, there is a wheel 
from which a weight $Z$ is suspended. Because of the collisions of gas molecules,
the vane is subjected to Brownian fluctuations. But
the ratchet is restricted to rotate in one direction only due to 
a pawl which in turn is connected to a spring.
Let ${\epsilon_{2}^{}}$ be the amount of energy
to overcome the elastic energy of the spring. Let in each step,
the wheel rotate an angle $\phi$ and the torque
induced by the weight be $Z$. Then the system requires a 
minimum of ${\epsilon_{1}^{}}={\epsilon_{2}^{}}+Z\phi$ energy to lift the weight
hanging from the axle.
Hence the rate of forward jumps of the ratchet is given as
$R_{\rm F}=r_0 e^{-{\epsilon_{1}^{}}/k_{\rm B}T_1}$,
where $r_0$ is a rate constant and $k_{\rm B}$ is Boltzmann's constant, 
which we set equal to unity. In other words, temperature has the dimensions of energy. 
A part of the energy ${\epsilon_{1}^{}}$ is converted into work $Z\phi$, 
and other is transferred as heat ${\epsilon_{2}^{}}$ to the cold thermal bath 
through the interaction between the ratchet and the pawl. 
Similarly, the rate of the backward jumps is $R_{\rm B}=r_0e^{-{\epsilon_{2}^{}}/T_2}$.
One may regard $Z\phi$ and $-Z\phi$ as the work done by and on the system, respectively. 
If $R_{\rm F} > R_{\rm B}$, this system works as two-reservoir 
heat engine. Then, the rates of heat related to the hot and the cold reservoirs, are  given as
\begin{equation}
\dot{Q}_1=r_0{\epsilon_{1}^{}} \left(e^{-{\epsilon_{1}^{}}/T_1}- 
e^{-{\epsilon_{2}^{}}/T_2}\right) >0, 
\label{q1dot}
\end{equation}
\begin{equation}
\dot{Q}_2=r_0{\epsilon_{2}^{}} \left(e^{-{\epsilon_{1}^{}}/T_1}- 
e^{-{\epsilon_{2}^{}}/T_2}\right) >0. 
\label{q2dot}
\end{equation}
According to the model, $\epsilon_{1}^{} > \epsilon_{2}^{}$, and so positivity of 
the fluxes implies: ${{\epsilon_{2}^{}}}/{T_2} > {{\epsilon_{1}^{}}}/{T_1}$. 
%
The power output, $P = \dot{Q}_1 - \dot{Q}_2$, is given by: 
\begin{equation}
P =  r_0({\epsilon_{1}^{}}  -  {\epsilon_{2}^{}})
\left(e^{-{\epsilon_{1}^{}}/T_1}- e^{-{\epsilon_{2}^{}}/T_2}\right).
\label{power}
\end{equation}
The efficiency of the engine, $\eta = {P}/{\dot{Q}_1}$ is given by 
\begin{eqnarray}
\eta = 1-\frac{{\epsilon_{2}^{}}}{{\epsilon_{1}^{}}} & \leq &  \eta_c. 
\label{efficiency-ratchet}
\end{eqnarray}
For given bath temperatures, 
it is natural to optimize the power output 
with respect to the internal energy scales 
${\epsilon_{1}^{}}$ and ${\epsilon_{2}^{}}$, 
which yields the following solution \cite{Tu2008}
\begin{eqnarray}
\epsilon_{1}^{*} &=& T_1 \left[{1 - (\eta_{c}^{-1}-1)\log(1-\eta_c)}\right],
\\
\epsilon_{2}^{*} &=& T_1 {(\eta_{c}^{-1}-1)(\eta_c - \log(1-\eta_c))},
\end{eqnarray}
with the expressions for the optimal power and EMP \cite{Tu2008} as given by
\begin{eqnarray}
P^* &=& {r_o e^{-1} T_1 \eta_c^2}{(1-\eta_c)^{(\eta_{c}^{-1}-1)}}, \label{powerTu} \\
\eta^* &=& {\eta_c}\left[{1-(\eta_{c}^{-1}-1)\log(1-\eta_c)} \right]^{-1}. \label{effTu}
\end{eqnarray}
Notably, $\eta^*$ depends only on the ratio of the reservoir temperatures. Further, 
the above expression of efficiency also holds for EMPs of a 
two-level atomic system \cite{WangTLS2013} and a simple model 
of classical particle transport \cite{Broeck2012}. Eq. (\ref{effTu}) 
has the following expansion for small values of $\eta_c$:
\begin{equation}
\eta^* = \frac{\eta_c}{2} + \frac{\eta_c^2}{8} + \frac{7\eta_c^3}{96} 
+ O(\eta_c^4). 
\label{compare1}
\end{equation}
The above series displays the universality up to second order mentioned
in the Introduction. 
\section{Ratchet in high temperatures regime}
In the following, we are interested in the regime, where the 
energies associated with forward and backward jumps are 
very small compared to the temperatures of reservoirs. Therefore, we can 
expand $e^{-{\epsilon_{1}^{}}/T_1}(e^{-{\epsilon_{2}^{}}/T_2})$ as 
Taylor series, say, up to first or second order. First, we look for a possible 
two-parameter power optimization in this regime.
Keeping terms up to the first order, we have the approximate expression 
for power as
\begin{equation}
P = r_0 ({\epsilon_{1}^{}}-{\epsilon_{2}^{}})\left(\frac{{\epsilon_{2}^{}}}{T_2} - 
\frac{{\epsilon_{1}^{}}}{T_1} \right).\label{powerfirst}
\end{equation}
We address the above approximation as the linear model \cite{Velasco}.
Similarities between the above model and a thermoelectric generator were 
recently discussed in Ref. \cite{Apertet2014}.

Now, a two-parameter optimization of the above expression, over 
$\epsilon_{1}^{}$ and $\epsilon_{2}^{}$,
yields the condition $T_1=T_2$, which is clearly not a meaningful result.
This suggests that the individual scales $\epsilon_1$ and $\epsilon_2$ may
not be varied independently within the linear model. We consider this 
idea in further detail in Sec. IV. However, if we retain terms up to 
second order in the exponentials, then power is given by 
\begin{equation}
P= r_0({\epsilon_{1}^{}}-{\epsilon_{2}^{}}) 
\left( \frac{{\epsilon_{2}^{}}}{T_2}-\frac{{\epsilon_{1}^{}}}{T_1}+
\frac{{\epsilon_{1}^{2}}}{2T_1^2}-\frac{{\epsilon_{2}^{2}}}{2T_2^2}\right).
\end{equation}
Now, optimizing the above expression over ${\epsilon_{1}^{}}$ and 
${\epsilon_{2}^{}}$, we get following solution
\begin{equation}
\epsilon_{1_{hot}}^* = T_1\frac{(4-3\eta_c)}{3(2-\eta_c)},  
\quad
\epsilon_{2_{hot}}^* = T_1\frac{(1-\eta_c)(4-\eta_c)}{3(2-\eta_c)}.
\label{e1e2}
\end{equation}
It is clear that the expression for efficiency remains as in 
Eq. (\ref{efficiency-ratchet}). Thus, we obtain the expressions for 
optimal power and EMP in the high temperatures regime
\begin{equation}
P_{hot}^* = \frac{2 r_o T_1\eta_c^2}{27(2-\eta_c)}, \label{Phot}
\end{equation}
\begin{equation}
\eta_{hot}^* = \frac{2-\eta_c}{4-3\eta_c}\eta_c.\label{zz}
\end{equation} 
If we expand $\eta_{hot}^*$ in Taylor series near equilibrium, we obtain 
\begin{equation}
\eta_{hot}^* = \frac{\eta_c}{2} + \frac{\eta_c^2}{8} + \frac{9\eta_c^3}{96} + O(\eta_c^4).
\label{compare2}
\end{equation}
The above series shows that the
universality of EMP up to second order \cite{Lindenberg2009,Tu2008}
survives in the high temperatures limit, using a non-linear approximation in the power output. 
The above form of efficiency is compared with  Eq. (\ref{effTu}) in figure 1, where 
we also compare the optimal power, Eq. (\ref{powerTu}), with the optimal power 
in high temperatures non-linear regime, Eq. (\ref{Phot}). It is to be noted that 
whereas the latter approximation overestimates EMP, the power output
is underestimated as compared to optimal power. 

%

\begin{figure}[ht]
 \begin{center}
\includegraphics[width=8.6cm]{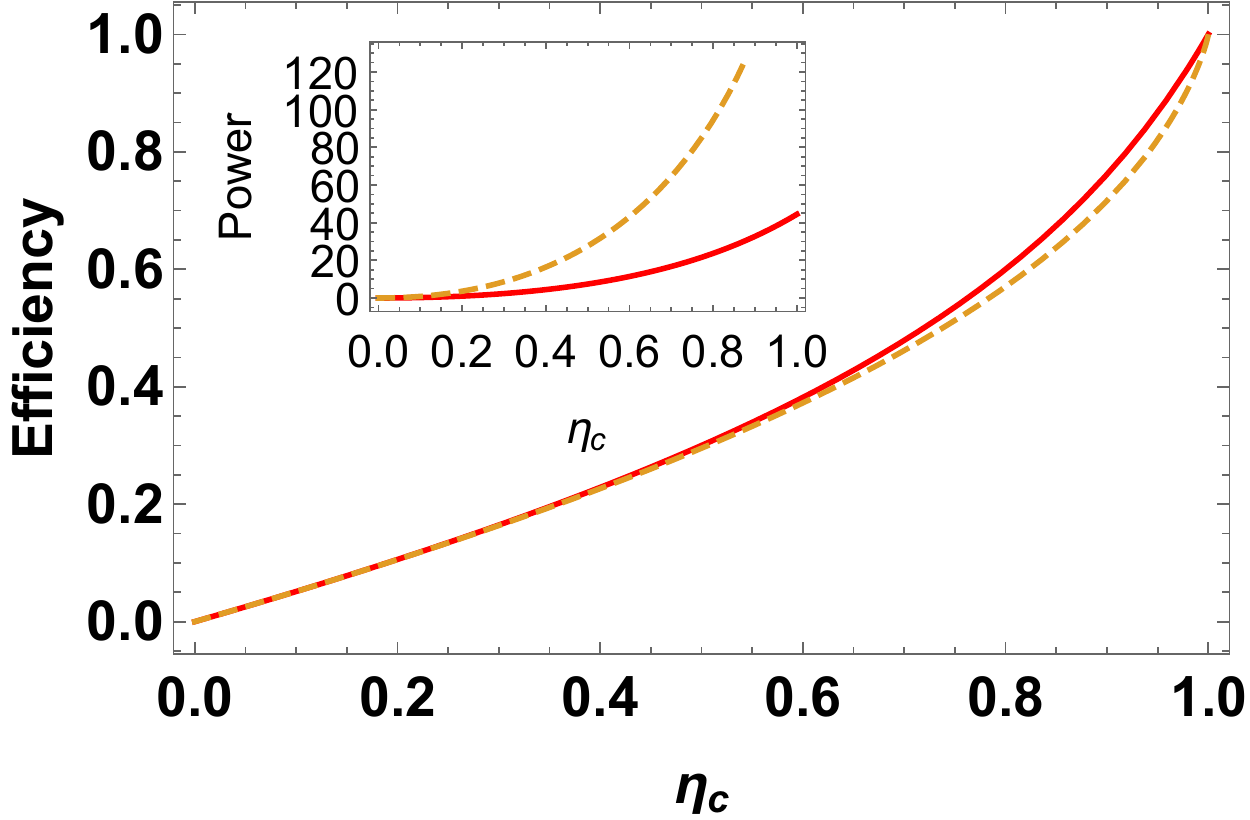}
 \end{center}
\caption{The efficiency of FS ratchet 
plotted against $\eta_c$.
Dashed curve and upper solid curves correspond to 
equations (\ref{effTu}) and (\ref{zz}) respectively, 
with the corresponding optimal power versus $\eta_c$, using $r_o=1$ and $T_1=600$.}
\end{figure}
\section{Linear regime with constraints}
In this section, we impose simple constraints on the energy scales of the 
ratchet system in the linear regime. This allows us to define a single-parameter 
optimization problem for power output, Eq. (\ref{powerfirst}). 
These constraints may be interpreted as a form of
control on the design of the device.
We are interested in the form of EMP under
the following constraints \cite{Kosloff2014}. 

(a) ${\epsilon_{1}^{}}= k_1>0$. Then optimizing power
(Eq. (\ref{powerfirst})) with respect to  
${\epsilon_{2}^{}}$, we get 
$\epsilon_2^* = {k_1}(2-\eta_c)/2$ and EMP as
\begin{equation}
\eta_{{\epsilon_{1}^{}}}= \frac{\eta_c}{2},
\label{eta1}
\end{equation}
a universal expression independent of the chosen $k_1$ value.  
\\
(b) On the other hand, consider setting ${\epsilon_{2}^{}}=k_2>0$. 
On optimization of power, we obtain 
$\epsilon_1^* = {k_2(2-\eta_c)}/{(2-2\eta_c)}$, and EMP as
\begin{equation}
\eta_{{\epsilon_{2}^{}}}= \frac{\eta_c}{2- \eta_c},
\label{eta2}
\end{equation}
which is again a universal formula depending only on the 
ratio of bath temperatures, but independent of the chosen 
constant $k_2$. Of course, the expressions for optimal 
power do depend on the chosen constant.
\\
(c) A more general constraint 
$\gamma{\epsilon_{1}^{}}+(1-\gamma){\epsilon_{2}^{}}=k_3$
where $0\leq \gamma \leq 1$. Here, the constraint involves
two fixed parameters. Optimization of power subject to 
this constraint, 
leads to the following optimal values:
\begin{equation}
\epsilon_1^* = \frac{k_3(2-(1-\gamma)\eta_c)}{2(1-(1-\gamma)\eta_c)},
\quad 
\epsilon_2^* = \frac{k_3(2-(2-\gamma)\eta_c)}{2(1-(1-\gamma)\eta_c)} 
\end{equation}
and the EMP is Schmiedl-Seifert (SS) efficiency \cite{Schmiedl2008}
\begin{equation}
\eta_{SS}= \frac{\eta_c}{2-(1-\gamma)\eta_c}.
\label{eta3}
\end{equation}
Clearly, (a) and (b) are special cases, with $\gamma=1$ and $\gamma=0$, respectively.
Here, EMP is independent of $k_3$, but depends on $\gamma$.
The above form has been obtained in Refs. \cite{Chen1989,Schmiedl2008,Esposito2010,Wangtu2012pre,JohalRai2016,Johal2017},
where the parameter $\gamma$ may be defined,  
for example, in terms of the ratio of the dissipation constants
or thermal conductivities of the thermal contacts \cite{Chen1989,Esposito2010}.
\\
(d) If the constraint ${\epsilon_{1}^{}}{\epsilon_{2}^{}} = k_4$ is imposed,
the optimal power is obtained at Curzon-Ahlborn (CA) efficiency \cite{Curzon1975}:
\begin{equation}
\eta_{CA}^{} = 1-\sqrt{1-\eta_c},
\end{equation}
at optimal values of $\epsilon_1$ and $\epsilon_2$:
\begin{equation}
\epsilon_1^* = \sqrt{k_4}(1-\eta_c)^{1/4},\quad \epsilon_2^* = \frac{\sqrt{k_4}}{(1-\eta_c)^{1/4}}.
\end{equation}
\section{Mapping to effective thermodynamic model} 
The expressions for EMP, obtained in the above, are also 
encountered in many thermodynamic models based on different
assumptions \cite{Chen1989,Schmiedl2008,Esposito2010,Wangtu2012pre,Johal2017}. 
They are obtained in finite-time 
as well as quasi-static models \cite{JohalRai2016,Johal2016} of heat engines.
Thus, it is natural to enquire about the thermodynamic
underpinning of the constrained FS model. 
In this section, we show that FS engine in the linear regime
can be mapped to a specific endoreversible model, under
the constraints described above. 
In the endoreversible approximation \cite{Curzon1975,Rubin,devos1985},
the work-extracting part of the engine operates in a reversible
way, and any irreversibility in the cycle is attributed solely
to thermal contacts with the reservoirs due to finite
conductance of the heat exchangers. 
\begin{figure}[ht]
 \begin{center}
\includegraphics[width=8.6cm]{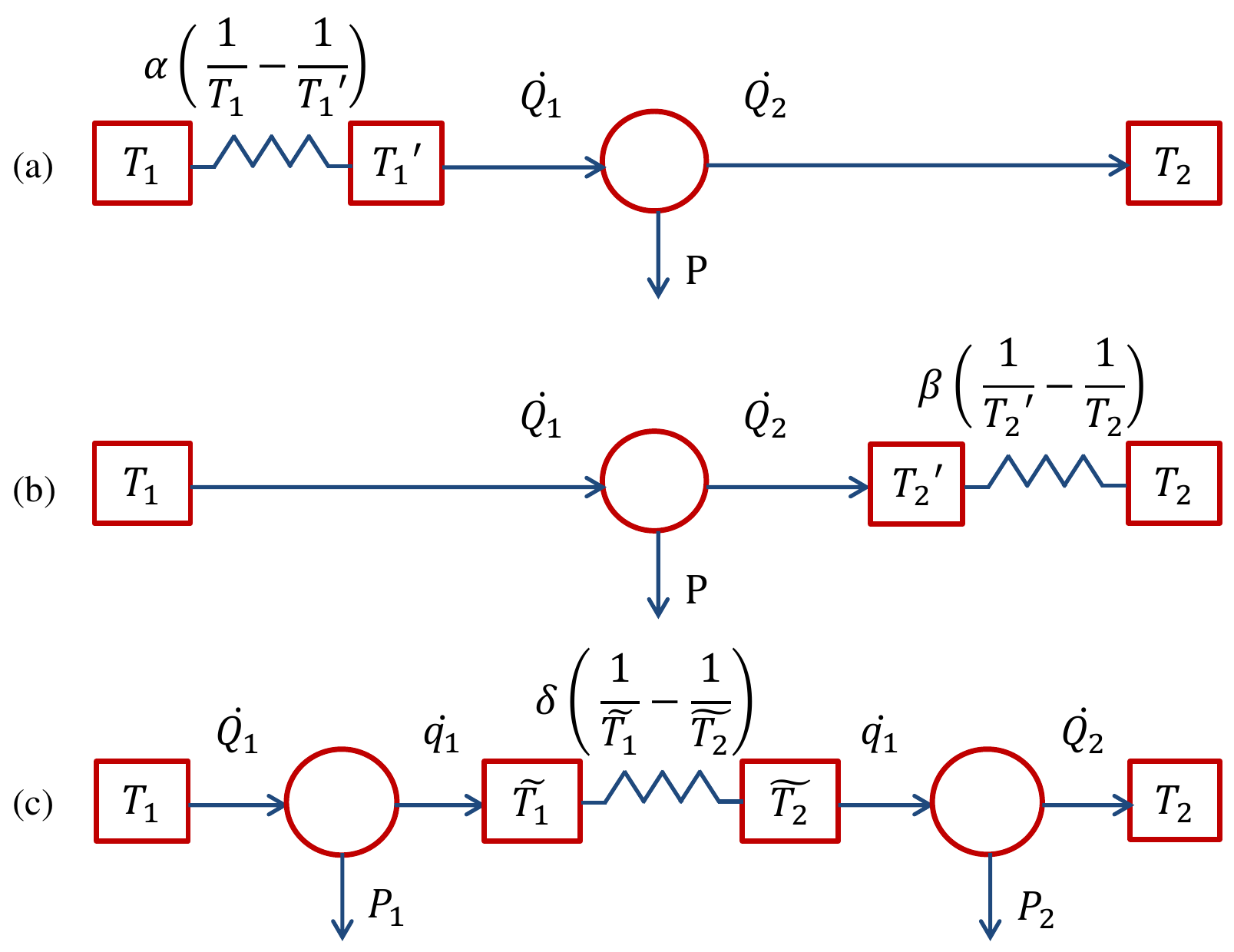}
\end{center}
\caption{Effective models of FS engine.
 Here $\alpha$, $\beta$ and $\delta$ are coefficients of heat conductance .
(a) Engine with resistance to incoming heat flux only; 
(b) Engine with resistance to outgoing heat flux only; 
(c) Two coupled reversible engines where the heat flow between the engines
    experiences resistance.}
\end{figure}
\par\noindent
(a) In the linear regime, 
the heat flux entering from the hot reservoir $\dot{Q}_1$, Eq. (\ref{q1dot}), 
is given by
\begin{eqnarray}
\dot{Q}_1 &=& r_0{\epsilon_{1}^{}}\left(\frac{{\epsilon_{2}^{}}}{T_2}-
\frac{{\epsilon_{1}^{}}}{T_1}\right). \\
 & \equiv & r_0{\epsilon_{1}^{2}} \left(\frac{1-\eta}{T_2} - \frac{1}{T_1}\right).
\end{eqnarray}
Here, we identify $T_1'=T_2/(1-\eta)$ as an effective temperature, satisfying $T_2 < T_1' < T_1$. 
Therefore, when we impose ${\epsilon_{1}^{}}=$ constant, the 
heat flux satisfies $\dot{Q}_1 \propto (1/T_1'-1/T_1)$, i.e. the flux 
is proportional to thermodynamic force as in linear irreversible thermodynamics. 
Then, it is assumed that the power is extracted between the
temperatures $T_1'$ and $T_2$, with reversible efficiency given by
$\eta = 1 - {T_2}/{T_1'}$.
%
%
Therefore,
\begin{equation}
 P = \eta\dot{Q}_1 = r_0\eta{\epsilon_{1}^{2}} 
\left(   \frac{1-\eta}{T_2} - \frac{1}{T_1}  \right).
\end{equation}
Optimizing power with respect to $\eta$ ($\partial P/\partial\eta=0$), we can obtain
EMP as in Eq. (\ref{eta1}).
%
\par\noindent
(b) Similarly, in terms of ${\epsilon_{2}^{}}$, the heat flux into the cold bath
can be written as
\begin{eqnarray}
\dot{Q}_2 &=& r_0{\epsilon_{2}^{2}} \left(\frac{1}{T_2}-\frac{1}{T_1(1-\eta)}\right)\nonumber
\\
 &\equiv &  r_0 {\epsilon_{2}^{2}} \left(  \frac{1}{T_2} - \frac{1}{T_2'}   \right)\nonumber,
\end{eqnarray}
where $T_2'=T_1(1-\eta)$ is an effective temperature, lying between values $T_1$ and $T_2$.
Thus, for a fixed value of ${\epsilon_{2}^{}}$, 
the heat flux $\dot{Q}_2$ is proportional to $(1/T_2'-1/T_2)$, which 
plays the role of thermodynamic force. 
In this case, power is extracted at Carnot efficiency between $T_1$ and $T_2'$: 
$\eta = 1 - {T_2'}/{T_1}$.
Therefore,
\begin{equation}
P  = \frac{\eta}{1-\eta} \dot{Q}_2 =
 r_0 {\epsilon_{2}^{2}}  \frac{\eta}{1-\eta}
\left( \frac{1}{T_2} - \frac{1}{T_1(1-\eta)}   \right).
\end{equation}
Optimizing the above equation with respect to $\eta$, we obtain Eq. (\ref{eta2}).

(c) For the linear constraint $\gamma{\epsilon_{1}^{}}+(1-\gamma){\epsilon_{2}^{}}=k_3$, 
the effective thermodynamic model is more interesting. 
In terms of $\eta$, this constraint equation can be written as
\begin{equation}
{\epsilon_{1}^{}}=\frac{k_3}{\gamma+(1-\gamma)(1-\eta)} \equiv \frac{k_3}{A}.
\label{defa}
\end{equation}
Then the expression for power becomes
\begin{eqnarray}
P &=&r_0 k_3^2\frac{ \eta }{A^2} \left( \frac{1-\eta}{T_2}-\frac{1}{T_1}\right),
\label{pwra}
  \end{eqnarray}
which can be  rewritten as follows:
\begin{equation}
P = r_0 k_3^2 \frac{\eta}{A} \left( \frac{1}{\tilde{T_2}} - \frac{1}{\tilde{T_1}} \right),
\label{pwra2}
\end{equation}
where the effective temperatures are defined as 
\begin{equation}
\tilde{T_1} = T_1 A, \qquad \tilde{T_2}=\frac{T_2 A}{1-\eta}.
\label{t12}
\end{equation}

Now we show that FS system in the linear regime, and under the general 
constraint, is equivalent 
to a system of two coupled Carnot engines in which heat flux 
leaving the first engine ($\dot{q}_1$), 
serves as input heat flux for the second engine,
through a finite heat conductance (see figure 2(c)). 
Thus consider the power output from the first engine:
\begin{equation}
P_1 = \frac{\eta_1}{1-\eta_1}\dot{q}_1, \label{power1}
\end{equation}
where $\eta_1$ is the reversible efficiency of engine 1
working between $T_1$ and $\tilde{T}_1$:
\begin{equation}
\eta_1 = 1-\frac{\tilde{T_1}}{T_1}=1-A,
\label{ef1}
\end{equation}
and 
\begin{equation}
\dot{q}_1=r_0 k_3^2\left(\frac{1}{\tilde{T_2}}-\frac{1}{\tilde{T_1}}\right), 
\end{equation}
is the heat flux leaving engine 1. Thus $r_0 k_3^2\equiv \delta$ is the heat transfer 
coefficient of the heat exchanger connecting engines 1 and 2.
So, we can rewrite Eq. (\ref{power1}) as 
\begin{equation}
P_1 = r_0 k_3^2 \frac{1-A}{A} \left( \frac{1}{\tilde{T_2}} 
- \frac{1}{\tilde{T_1}} \right)\label{power11}.
\end{equation}
Now, engine 2 operates at Carnot efficiency 
$\eta_2$ between temperatures $\tilde{T_2}$ and $T_2$:
\begin{equation}
\eta_2 = 1-\frac{T_2}{\tilde{T_2}} = 1-\frac{1-\eta}{A},
\label{ef2}
\end{equation}
with the input heat flux as $\dot{q}_1$.
Hence, power of engine 2, $P_2=\eta_2 \dot{q}_1$  can be written as
\begin{equation}
P_2 = r_0 k_3^2 \left(1-\frac{1-\eta}{A}\right)
\left( \frac{1}{\tilde{T_2}} - \frac{1}{\tilde{T_1}} \right).
\label{power22}
\end{equation}
Adding equations (\ref{power11}) and (\ref{power22}), we get
\begin{equation}
P_1 + P_2 =  P,
\end{equation}
which is the total power, Eq. (\ref{pwra2}). 
Alternately, we can write  $P_1 = (1-\gamma)P$ and $P_2 = \gamma P$.
Optimizing $P$ with respect to $\eta$, we obtain 
Eq. (\ref{eta3}). It is clear that the maximum of $P_1$ and $P_2$ 
is also reached at the same value of $\eta$ as of $P$. Thus optimality of
$P$ for the overall engine implies optimal power output 
of the sub-engines.

Now, the values $\gamma =0$ and $\gamma =1$ 
correspond to the special cases (a) and (b), respectively.  
Using $A=\gamma+(1-\gamma)(1-\eta)$, we can write
\begin{equation}
\eta_1^{} = (1-\gamma)\eta,\qquad \eta_2^{} = \frac{\gamma \eta}{1-\eta+\eta\gamma}.
\end{equation}
Also, the manner in which the two sub-engines are coupled, implies that 
the efficiencies of the sub-engines are related to the 
overall efficiency as: $\eta = 1- (1-\eta_1)(1-\eta_2)$.

Finally, using Eq. (\ref{eta3}), the EMPs for 
engine 1 and 2 are given by
\begin{equation}
\eta_1^* = \frac{(1-\gamma)\eta_c}{2-(1-\gamma)\eta_c},\qquad \eta_2^* = 
\frac{\gamma\eta_c}{2-2(1-\gamma)\eta_c}.
\end{equation}
\par\noindent
(d) For the constraint ${\epsilon_{1}^{}}{\epsilon_{2}^{}}=k_4$, 
Eq. (\ref{powerfirst}) for power becomes simplified as
\begin{eqnarray}
P &=& r_0 k_4 \eta  \left(  \frac{1}{T_2} - \frac{1}{T_1(1-\eta)} \right) . \label{Pcouple} \\
 & \equiv & r_0 k_4  \frac{\eta }{\sqrt{1-\eta}} \left( \frac{1}{\bar{T_2}} 
 - \frac{1}{\bar{T_1}} \right ), \label{Pcouple2}
\end{eqnarray}
where we have defined 
\begin{equation}
\bar{T_1} =  T_1\sqrt{1-\eta}, \qquad \bar{T_2} = \frac{T_2}{\sqrt{1-\eta}},
\end{equation}
as the effective temperatures. Further,
it is useful to decompose $\eta/\sqrt{1-\eta}$ as follows:
\begin{equation}
\frac{\eta}{\sqrt{1-\eta}} = \frac{1-(1-\eta)}{\sqrt{1-\eta}} =
\frac{1}{\sqrt{1-\eta}} -\sqrt{1-\eta}.
\end{equation}
Thus, we can express Eq. (\ref{Pcouple2}) in the following form
\begin{equation}
P =  r_0 k_4 \left( \frac{1}{\sqrt{1-\eta}} - 1 \right)   
\left( \frac{1}{\bar{T_2}} - \frac{1}{\bar{T_1}} \right ) 
  + r_0 k_4 (1-\sqrt{1-\eta}) \left( \frac{1}{\bar{T_2}} 
 - \frac{1}{\bar{T_1}} \right ),    
 \end{equation}
which can be rewritten as
\be
 P=\frac{\eta_1'}{1- \eta_1'}  \dot{q}_1' +  \eta_2'\,\dot{q}_1' \equiv P_1 + P_2,
 \ee
where $\eta_1'(\eta_2')$  is Carnot efficiency of engine 1(2) 
operating between the temperatures $T_1(\bar{T_2})$ and $\bar{T_1}(T_2)$, defined as
\begin{eqnarray}
\eta_1' &=& 1 - \frac{\bar{T_1}}{T_1} = 1-\sqrt{1-\eta},\label{same1}
\\
\eta_2' 
 &=& 1-\frac{T_2}{\bar{T_2}} = 1-\sqrt{1-\eta}, 
\label{same2} 
\end{eqnarray}
and
\be
\dot{q}_1' = r_0 k_4 \left( \frac{1}{\bar{T_2}} - \frac{1}{\bar{T_1}} \right)
\ee
is the heat flux leaving engine 1 and entering engine 2 (see figure 3(a)). Here $r_0 k_4$ is 
heat transfer coefficient of the heat exchanger connecting engines 1 and 2.
We note that engine 1 and engine 2 deliver power at same efficiency.
%
Then, the expressions for 
$\eta_1'$ and $\eta_2'$, at optimal power, are given by:
\begin{equation}
\eta_1'^* = \eta_2'^* = 1 - (1-\eta_c)^{1/4}.
\end{equation}
\begin{figure}[ht]
 \begin{center}
\includegraphics[width=8.6cm]{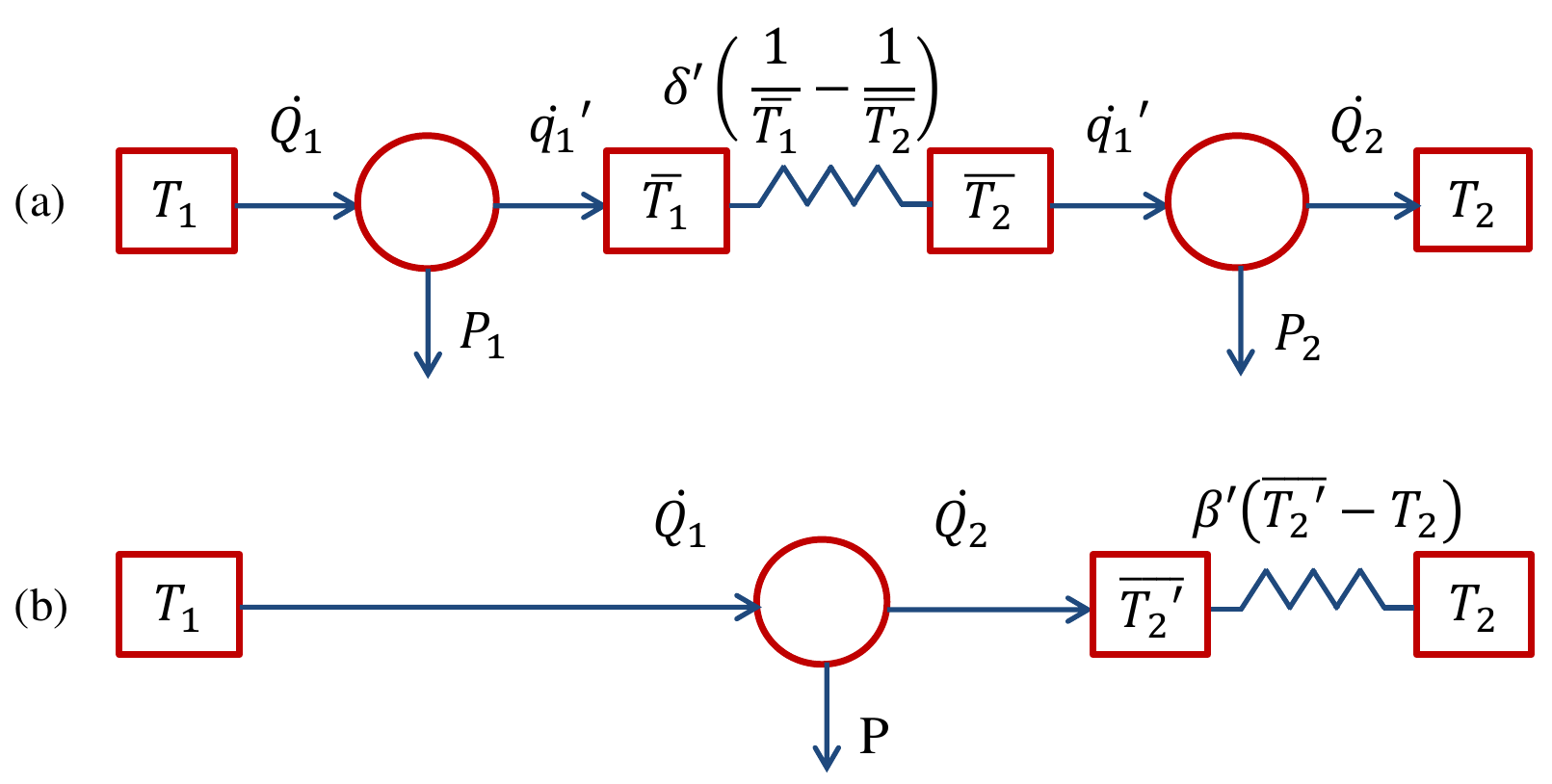}
\end{center}
\caption{Two different thermodynamic models of FS ratchet for the constraint $\epsilon_1\epsilon_2=k_4$. 
(a) Two Carnot engines connected by thermal conductance $\delta' = r_0 k_4$. 
(b) Carnot engine subjected to finite thermal conductance $\beta'$ for the outgoing heat flux obeying 
Newton's law.}
\end{figure}
\section{Discussion and Summary}
%
%
Our choice of constraints is motivated by the fact that 
both $\epsilon_1$ and $\epsilon_2$ are the control 
parameters of the ratchet system. It is possible to tune 
either of them to obtain a desired performance of 
the engine. In other words, energy constraints can be imposed by setting
a design goal. On the other hand, it is not straightforward to appreciate 
the nature of control with the general linear constraint (c), though one can consider 
the equivalent thermodynamic model with effective 
temperatures as in Eq. (\ref{t12}). For a given value of $\gamma$, we can 
tune these temperatures and thus the efficiencies of engines 1 and 2.
For $\eta=0$, we have $\tilde{T_1}=T_1$
and $\tilde{T_2}=T_2$. In the reversible limit, 
when $\eta=\eta_c$, we have $\tilde{T_1}=\tilde{T_2}=
\gamma T_1+(1-\gamma) T_2$, see figure 4.  From Eqs. (\ref{power11}) and 
(\ref{power22}), it is also clear that the power vanishes as 
$\tilde{T_1}\rightarrow \tilde{T_2}$.
\begin{figure}[ht]
 \begin{center}
\includegraphics[width=8.6cm]{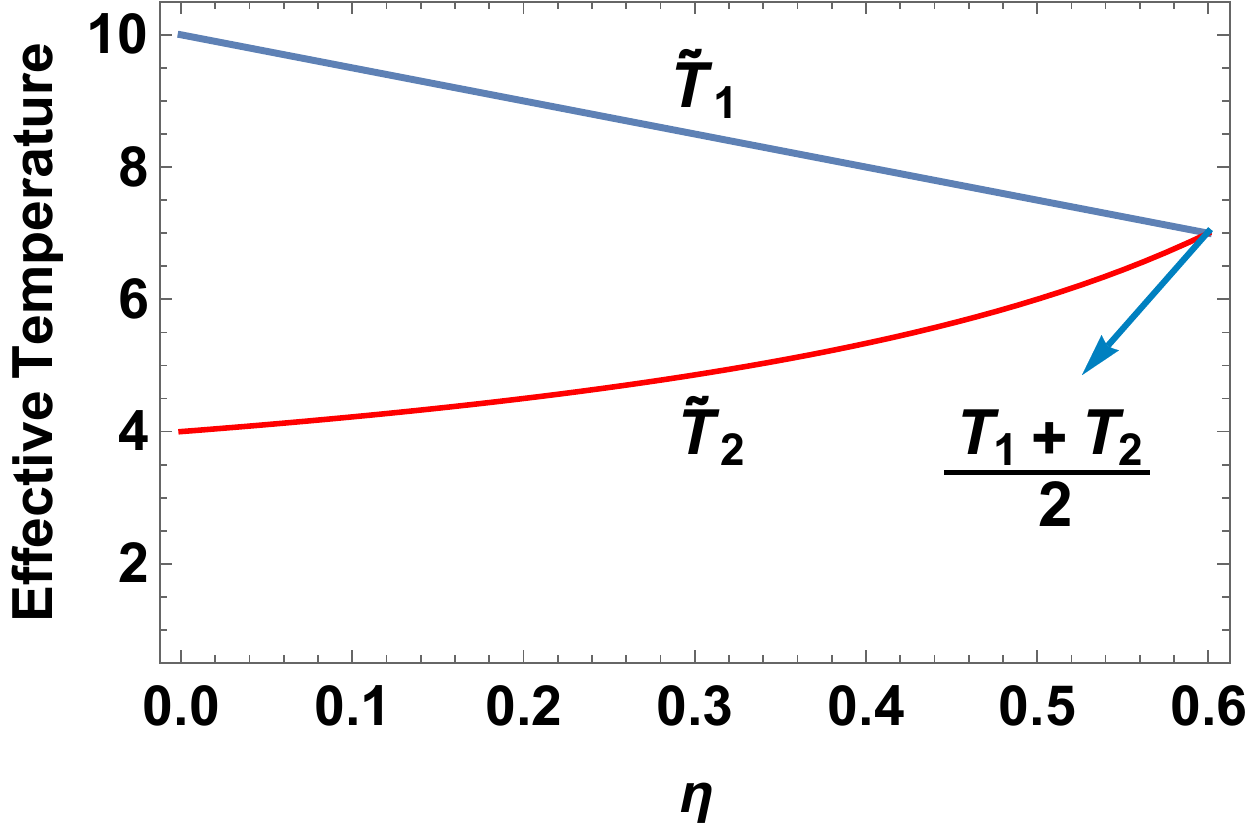}
 \end{center}
 \caption{Effective temperatures $\tilde{T_1}$($\tilde{T_2}$)
 with constraint (c), as plotted against the 
 efficiency of the ratchet engine. Here $T_1=10$, $T_2=4$ and $\eta_c=0.6$. The upper curve 
 represents the effective temperature of engine 1 and the 
 lower curve represents the effective temperature of engine 2.}
\end{figure}
Similar considerations can be made regarding the control of the effective
temperatures in the case of constraint (d).

However, note that the proposed thermodynamic model for the constrained FS system may
not be unique. This may be shown by considering the case (d). 
We have mapped this model to two coupled reversible engines
connected by a heat flow with an inverse-temperature law. 
It has been shown that the EMP in this model is CA-efficiency.
Usually, CA-value is associated with EMP for endoreversible
models with Newtonian heat flows, i.e. heat flux is proportional
to the difference of temperatures between which the heat flow
takes place \cite{Curzon1975,devos1985}. In fact, it is possible to imagine  
an alternate model as follows (see figure 3(b)).
By rewriting the power output, we  get 
\begin{eqnarray} 
P&=& \frac{r_0 k_4}{T_1 T_2} \frac{\eta}{1-\eta} \left(T_1(1-\eta)-T_2 \right)\nonumber
\\
& \equiv & k'  \frac{\eta}{1-\eta} (\bar{T_2'}-T_2),
\end{eqnarray}
where we define $\bar{T_2'} = T_1(1-\eta)$ as the effective temperature and 
$k'=r_0 k_4/T_1 T_2$ as the coefficient of the exiting heat flux 
$\dot{Q}_2 =k'(\bar{T_2'}-T_2)$, between temperatures $\bar{T_2'}$ and $T_2$. 

Concluding, we have considered the optimization of output power in FS ratchet
in the high temperatures regime, when the internal energy
scales are much smaller in comparison to bath temperatures. 
A two-parameter optimization is possible if one includes
the quadratic terms in the expansion of the exponentials.
 For the linear model,
we have considered simple constraints on the internal scales, 
and obtained some well-known forms of EMP, such as SS-efficiency and 
CA-efficiency. The reason for these similarities
is appreciated by showing that  the constrained FS system 
can be mapped to a finite-time endoreversible model
with appropriately defined heat flows, using effective temperatures.
Finally, due to a formal analogy between FS system
in the linear regime and thermoelectric models \cite{Apertet2014},
and also specific types of quantum heat 
engines in the hot temperatures regime \cite{Kosloff2014}, 
the present analysis can provide a useful 
perspective on a broader class of energy conversion systems.  

\section*{References}
\bibliography{biblofeynman}

\end{document}